\documentclass{article}

\usepackage{arxiv}

\usepackage[utf8]{inputenc} 
\usepackage[T1]{fontenc}    
\usepackage{hyperref}       
\usepackage{url}            
\usepackage{booktabs}       
\usepackage{amsfonts}       
\usepackage{nicefrac}       
\usepackage{microtype}      
\usepackage{cleveref}       
\usepackage{lipsum}         
\usepackage{graphicx}
\usepackage{doi}

\usepackage{multicol}

\usepackage{tabularx}
\usepackage{capt-of}

\title{Machine Learning and Port Scans: A Systematic Review}

\author{
    Jason M. Pittman \\
    ORCID: 0000-0002-5198-8157 \\
}

\hypersetup{
    pdftitle={Machine Learning and Port Scans: A Systematic Review}
    pdfsubject={}
    pdfauthor={Jason M. Pittman}
    pdfkeywords={systematic review, machine learning, port scanning, cybersecurity, algorithms, training data}
}

\begin{document}
\maketitle

\begin{abstract}

Port scanning is the process of attempting to connect to various network ports on a computing endpoint to determine which ports are open and which services are running on them. It is a common method used by hackers to identify vulnerabilities in a network or system. By determining which ports are open, an attacker can identify which services and applications are running on a device and potentially exploit any known vulnerabilities in those services. Consequently, it is important to detect port scanning because it is often the first step in a cyber attack. By identifying port scanning attempts, cybersecurity professionals can take proactive measures to protect the systems and networks before an attacker has a chance to exploit any vulnerabilities. Against this background, researchers have worked for over a decade to develop robust  methods to detect port scanning. While there have been various surveys, none have focused solely on machine learning based detection schemes specific to port scans. Accordingly, we provide a systematic review of 15 papers published between February 2021 and January 2023. We extract critical information such as training dataset, algorithm used, technique, and model accuracy. We also collect unresolved challenges and ideas for future work. The outcomes are significant for researchers looking to step off from the latest work and for practitioners interested in novel mechanisms to detect the early stages of cyber attack.   

\end{abstract}


\begin{multicols}{2}

\section{Introduction}

Cybersecurity incidents continue to plague digital life. While a significant portion of incidents result from phishing and malware, 45\% are the result of network-based cyber attacks \cite{statistia}. These cyber attacks follow a pattern or procedure. Existing models and methodologies vary in the number of steps. However, the first step is universally understood to be \textit{reconnaissance}. In turn, reconnaissance most often includes some type of port scanning.

Port scanning is a technique to enumerate target endpoints. Confusingly, port scanning can be both a legitimate engagement \cite{Wang2014} or a malicious precursor to escalating intrusion  \cite{Baig2007, yadav2015technical}. A general issue is differentiating between what may be an authorized benign instance of host enumeration and a malicious scanning of active hosts and their available ports. Furthermore, if we accept port scanning as a necessary prelude to cyber attack, then we want to develop a means to detect port scanning with high certainty. To this end, there is a small but growing literature on detecting port scanning. The literature ranges from early intrusion detection mechanisms \cite{heberlein1990network} to sophisticated machine learning techniques \cite{henry2023composition}. There have been several comparative surveys during this time, most recently Aamir et al. \cite{Aamir2021} and \cite{krishna2021}. However, there has not been a systematic review of the literature.

Literature reviews are invaluable to a field of study. Reviews provide an understanding of the existing research by establishing a foundation of knowledge. Reviews also clarify existing knowledge related to a given problem. Both functions guide new investigations and reduce overlap or unnecessary duplication of work. Yet, new reviews are necessary as the field grows, new techniques are discovered, and new technologies are released which impact forms of inquiry. Accordingly, the purpose of this study is to provide a systematic review of existing literature using machine learning algorithms to detect port scanning.  

The remainder of this work is organized in a way which (a) situates the systematic review in existing knowledge and (b) maximizes understanding of the cutting edge. The first is achieved by discussing port scanning and detection of such port scanning literature. Thereafter, we present the research method and techniques used to find, organize, and analyze research published since 2021. Finally, we demonstrate the findings of the analysis in terms of quantitative results from the existing research. 
 
\section{Related Work}

The work most proximal to this study exists in two categories: scanning TCP/IP ports and detection of those scans. The following discussion is not intended to be exhaustive. Rather, we offer background research that we view as seminal and salient.

\subsection{Port Scanning}

Port scanning uses features of TCP/IP to enumerate computing systems on a network. As different network protocols use different ports, it's essential to scan a wide range of ports to gather complete information. This is because vulnerabilities can exist in all protocols. The total number of ports that can be scanned is 65535, with ports 0 to 1023 being well-known, ports 1024 to 49151 being registered, and ports 49152 to 65535 being dynamic or private.

The origin of the phrase \textit{port scanning} in the academic literature can be traced back to the early days of computer networking. In the late 1980s and early 1990s, as the Internet was growing and becoming more widely used, there was an increasing need for tools to help network administrators and security professionals understand the state of their networks. One of the key tasks for these professionals was identifying which network services were running on which hosts, and which ports on those hosts were open or closed. This process became known as \textit{port scanning}.

One of the earliest references to port scanning in the literature is found in Fyodor \cite{lyon1998}, which described a method for determining the operating system of a remote host by sending probes to specific ports and analyzing the responses. The work explained the operating system of a host can be determined by analyzing the TCP/IP stack's behavior and its responses to different types of probes, such as the initial sequence numbers (ISNs) and the options in the TCP headers of the responses. Further, the paper also described how to use this technique to fingerprint the operating system of a remote host as well as the limitations and challenges of the technique. Additionally, the paper introduced the first version of an open-source tool named \textit{nmap} (Network Mapper) that implements this technique for Remote OS detection. While nmap is not the only port scanner available, it is featured heavily throughout the literature.

De Vivo et al. \cite{de1999review} generalizes from the port scanning foundation provided in Fyodor \cite{lyon1998} and several \cite{comer1994probing, wright1995tcp} others. The significance of De Vivo et al. \cite{de1999review} emerges from the rigorous classification applied to port scanning techniques and procedures. The paper described the different types of port scans, such as TCP connect scans and SYN scans as \textit{classical}. This is in relation to \textit{indirect} and \textit{stealth} scanning. The latter is also referred to as a FIN, XMAS, or NULL scan. The former is realized by bouncing scans off of a zombie endpoint. The work goes on to describe scanning techniques. These includes \textit{decoy} scanning, \textit{fragmented} scanning, and \textit{coordinated} or \textit{distributed} scanning, UDP scanning, and ICMP sweeping. 

Barnett et al. \cite{barnett2008towards} presented a classification system for network scanning techniques. The significance of the work is in establishing a clear and organized classification of the different types of network scanning techniques that exist and their use cases. To that end, the authors propose a taxonomy categorizing network scanning techniques based on the level of interaction with the target system, the type of information gathered, and the purpose of the scan. This extends De Vivo et al. \cite{de1999review} in both types and techniques.

Barnett et al. \cite{barnett2008towards} add two additional network scanning techniques to the three presented by De Vivo et al. These are \textit{vertical}, \textit{horizontal}. Further, Barnett et al. differentiate between OSI Model layer 2 scans and layer 3 scans with overlaying attributes according to speed (slow, medium, rapid) and distribution (one-to-one, one-to-many, many-to-one, many-to-many). Barnett et al. also describe how scan types from prior work (e.g., De Vivo et al.) map to their categories. The mapping is more pronounced when attributes encompassing speed and distribution were considered.   

Bou et al. \cite{bou2013cyber} demonstrated a comprehensive overview of the different types of cyber scanning techniques that are used to identify various features of networks. The authors divide port scanning techniques into two main categories: \textit{passive} and \textit{active}. Passive scanning techniques involve listening to network traffic to gather information about the target network without sending any packets. Active scanning techniques involve sending packets to a target host to elicit a response, which can be used to determine the host's characteristics and identify vulnerabilities. 

The work extends the categorization by describing different techniques of passive and active scanning techniques. This calls back to the organizational structure provided by De Vivo et al. \cite{de1999review} and Barnett et al. \cite{barnett2008towards} but differs in semantics. For instance, the De Vivo et al. classical, indirect, and stealth scans map under the nature of active and passive scanning. Further, the semantic developed by \cite{barnett2008towards} around relations between scanner and target (e.g., one-to-many) falls under \textit{approarch} in Bou et al. \cite{bou2013cyber}.  Bou et al. also offered \textit{strategy} as a way to categorize directional relationship between scanner and target. 

Roy et al. \cite{roy2022survey} claimed a gap exists in the literature on classifying and categorizing adversarial reconnaissance processes. The claim stands to reason given the authors first delineate between \textit{technical} and \textit{non-technical} reconnaissance techniques. Technical reconnaissance included network scanning, or \textit{cyber scanning} as Roy et al. refer to it, as a remote technique. The authors then differentiate between \textit{host detection} and \textit{port enumeration} which stand as a combined label for all of the scanning techniques outlined by De Vivo et al \cite{de1999review} (i.e., ICMP, SYN, Full Connect, etc.). 

While Roy et al. \cite{roy2022survey} did not add anything new to the port scanning taxonomy, the authors did connect the prior research by De Vivo et al. \cite{de1999review} and Barnett et al. \cite{barnett2008towards} to a burgeoning literature around detection of port scanning. 

\subsection{Detecting Port Scanning}
Port scanning, as a reconnaissance technique, is detectable. ML is a compelling solution to detecting otherwise undetectable port scans because of its ability to correlate seemingly unrelated features across enormous datasets. Yet, not all ML algorithms work in the same way or have the capability to address the same problem. Furthermore, there are a variety of ML algorithms types- classification, regression, deep learning, and so forth- with a diversity of implementation variations. 

The majority of works investigating ML for detecting port scanning over the past decade and a half include at a review of prior ML algorithm performance. Such work exists as a quasi review with an add-on quantitative analysis of an algorithm not featured in the prior research. We use the term \textit{quasi} because these works review algorithms by running each against a common training and evaluation dataset. These studies do not rely on results from the prior research responsible for introducing the algorithm to the field. This is close to the notion of a meta-analysis but not precisely so. Still, the quasi reviews are particularly significant for researchers and practitioners looking to get up to speed on the state of the field in short order. We discovered two such works and include those as foundational literature which we directly extend with our systematic review. 

Aamir et al. \cite{Aamir2021} investigated the detection of characteristics of port scanning and analyzed the performance of 22 ML algorithms. The algorithms included decision trees, discriminant analysis, support vector machines (SVM), k-nearest neighbors (KNN), and ensemble classifiers. The authors used the CICIDS2017 dataset with a 70\%training and 30\% evaluation split. Of the 22 ML algorithms examined, nine demonstrated more than 85\% classification (testing) accuracy. Specifically, Aamir et al. identified Fine Gaussian SVM as best performing algorithm with 99\% testing and 99\% training accuracy. False negative rates are provided for all 22 algorithm experiments. Further, for fast training with high accuracy scores, discriminant analysis was more accurate and efficient in classifying port scans. Aamir et al. did not discuss the type of port scans detected nor what scanning techniques were present in the dataset. 

Krishna et al. \cite{krishna2021} also investigated port scan detection using a variety of ML algorithms. The authors analyzed fewer algorithms but used the same training and evaluation dataset as Aamir et al. \cite{Aamir2021}. Krishna et al. examined two of the algorithms as Aamir et al., SVM and decision trees. Krishna et al. also evaluated random forest and logistic regression. Unlike any other study we found in the literature, Krishna et al. do not present results. Instead, the authors include snapshots of their Juypter notebook as figures. On one hand, including code makes the work repeatable and reproducible. On the other hand, one would need to repeat and reproduce the study to obtain algorithm performance values.

Both Aamir et al. \cite{Aamir2021} and Krishna et al. \cite{krishna2021} represent the predominant type of research in the literature. That is, existing port scan detection research frequently examines ML algorithm performance by direct experimentation rather than reference to prior experimentation. Consequently, the findings from these studies are scattered throughout the literature. Anyone interested in extending the field is left to trace through the forest to find relevant trees. With that in mind, the goal of this work was to systematically review results published since 2021 to catalog ML algorithm performance in detecting port scans.

\section{Method}
This work employed a systematic literature review methodology. Systematic literature review is a well-defined method that is used to identify, evaluate, and interpret all of the available research on a particular topic \cite{cooper1998synthesizing, kitchenham2004procedures}. The process is designed to be comprehensive, unbiased, and transparent, and it involves a number of steps, including formulating a research question, searching for relevant literature, selecting studies, extracting data, and synthesizing the results \cite{cooper1998synthesizing, petticrew2008systematic}. Systematic literature reviews are increasingly used in the field of software engineering and other technical fields, but also in other scientific fields, as a way to provide an in-depth understanding of existing knowledge on a topic.

A systematic literature review differs from other literature review methods in several ways. A traditional literature review, also called a narrative review \cite{cooper1998synthesizing}, is typically less structured and less systematic. It is often used to provide an overview of the current state of knowledge on a topic, but it is not as rigorous as an SLR in terms of the search and selection process.

With the design of systematic reviews in mind, we pose four questions.

RQ1: What machine learning algorithms have been used to detect port scanning?

RQ2: What were the detection rates and false positive rates for those algorithms?

RQ3: What datasets were used for training and evaluation of those algorithms?

RQ4: What port scanning types and techniques were used for evaluation of those algorithms?

We constrained the literature search to 2021 and newer. We did so based on the last relevant reviews being published in 2021 \cite{Aamir2021, krishna2021}. While we recognize the majority of work in detecting port scans exists between 2010 and 2021, research is still progressing in this research area and a systematic review of published research since 2021 holds significance for researchers and practitioners alike.

No literature search will produce perfect results. However, careful attention to search strings can yield sufficient results so as to be thorough. We used \textit{"detecting port scan" AND "machine learning"} as a starting search string. The search returned 61 papers. For comparison, searching with \textit{"port scan" AND "machine learning"} produced 952 results. Meanwhile, \textit{"detect port scan" AND "machine learning"} produced 21 results. Often researchers use the commonly accepted short form of machine learning, ML. Thus, we used \textit{"ML" AND "detecting port scan"} to cross-check the search. This produced 43 results. The variant of \textit{"detect port scan" AND "ML"} found 12 papers. 

We manually reviewed each work to ensure each study included detection of port scanning. Manual review was necessary because some work folds port scan detection into an overarching intrusion detection framework. We were left with 15 studies as our dataset after this step.

Data extraction from the selected papers is an important step to properly answer the research questions. In this study, we used the following data form to extract the needed information: (a) year of publication; (b) authors; (c) source of publication; (d) citation count; results (accuracy as F1); (e) dataset source; and (f) algorithms as task, technique, and procedure (TTP). We also included the port scanning types and techniques when such were available in the research. 

\section{Results}
We separate the results of the systematic review into two sections. The first section provides an overview of our dataset. We describe the literature features for ease of future reference. Then, we present a breakdown of that literature by algorithm. As with Aamir et al. \cite{Aamir2021} and Krishna et al. \cite{krishna2021}, some work experimented with more than one ML algorithm. Such research appears in multiple categories below.

\subsection{The Literature}
We analyzed 15 studies published since 2021 (Table 1). Seven studies were from 2021, six were from 2022, and a single study appeared in early 2023. The remaining study was from 2020 which we included as a specific exception. This is discussed in the Neural Network (NN) algorithm section below. 

The sample encompassed six total ML algorithms. The majority (10) of studies examined a single ML algorithm while five studies examined more than one. The literature published in 2021 spanned all six algorithms whereas literature from 2022 focused on a single algorithm (with one exception). Random Forest (RF) and SVM were the most investigated algorithms in the 2021 subset. A variety of NN implementations appeared throughout the 2022 subset. Nature-inspired (NI) appeared once while Regression (R) and Naive Bayes (NB) were studied three and five times respectively.

Six studies have not been cited. Six studies have been cited more than once with 25 being the highest citation count. Only one study \cite{al2021detecting} included a paper \cite{Aamir2021} from the literature population in its related work. The other 14 papers exist independent of one another with only indirect relations from support research in general ML or cybersecurity.

\begin{center}
\captionof{table}{Literature using ML to detect port scans}
\begin{tabular}{cccc}
	\textbf{Authors} & \textbf{Year} & \textbf{Cited} & \textbf{TTP} \\ \hline \hline
	Hartpence et al. & 2020 & 6 & NN\\
	Algaolahi et al. & 2021 & 1 & RF,SVM\\
	Baah et al. & 2021 & 0 & RF,SVM,NB\\
	Sirisha et al. &	2021 & 4 & RF,R,NB\\
	Liu et al. & 2021 & 21 & NI\\
	Bertoli et al. &	2021 & 25 & RF,SVM,R,NB,NN\\
	Mohseni et al. & 2021 & 1 & RF\\
	Al-Haija et al. & 2021 & 9 & NB\\
	Bakaletz & 2022 & 0 & NB\\
	Tojeiro et al. & 2022	 & 0 & R\\
	Singh et al. & 2022 & 2 & NN\\
	Lv et al. & 2022	 & 0 & NN\\
	Kirtas et al. & 2022 & 1 & NN\\
	SaiKiran et al. & 2022 & 0 & RF,SVM,NN\\
	Henry et al. & 2023 & 0 & NN\\ \hline
\end{tabular}
\end{center}

\subsection{The Algorithms}
We found six machine learning algorithms in the literature sample. The following sections present a summary for each algorithm and the relative meaning of using it to detect port scanning. We summarize each algorithms's performance in terms of accuracy and false positives. We also present the dataset used to train and evaluate the models when such are revealed in the source literature.  

\subsubsection{Random Forest}
Random Forest is good for classification problems, particularly in cases where there are many features and interactions among features. The algorithms is also useful for feature selection and handles missing data well.

Six studies out of the 15 study sample experimented with the RF algorithm \cite{algaolahi2021port, bertoli2021end, sirisha2021intrusion, mohseni2021density, baah2022enhancing,  saikira2022detection}. Algorithm performance ranged from 78.09\% to 100\% across those studies. One paper \cite{bertoli2021end} included source code or a link to a source code repository (e.g., GitHub). Four different datasets were used, three of which do not appear in other algorithm categories.

Two studies discussed the types of port scans present in training and evaluation data. SaiKiran et al. \cite{saikira2022detection} mentioned \textit{port sweep} but did not specify further. Bertoli et al. \cite{bertoli2021end} conducted training and evaluating against the full spectrum of port scan types. Further, the authors included port scan data from five different port scan tools.

\begin{center}
\captionof{table}{Random Forest Algorithm Performance}
\begin{tabular}{cccc}
	\textbf{Authors} & \textbf{Accuracy (F1)} & \textbf{Dataset} \\ \hline \hline
	Algaolahi et al. & 99.75 & CICIDS2017\\
	Baah et al. & 99.98 & CICIDS2017\\
	Sirisha et al. & 78.09 & NSLKDD \\
	Sirisha et al. & 84.14 & CICIDS2017 \\
	SaiKiran et al. & 99.93 & CICIDS2017 \\
	Mohseni et al. & 99.94 & CICIDS2017 \\
	Bertoli et al. & 96.00 & MAWILab \\ 
	Bertoli et al. & 100.00 & Bonafide \\ \hline
\end{tabular}
\end{center}

\subsubsection{Support Vector Machine (SVM)}
Support Vector Machine (SVM) is good for classification and regression problems, especially in cases where the data has clear boundaries and is not noisy. It works well for datasets with a limited number of features.

SVM is different from Random Forest in that it uses a boundary (a hyperplane) to separate the data into classes, whereas Random Forest creates multiple decision trees and aggregates their predictions to make a final decision. SVM is best suited for cases where the boundary between classes is well defined and clear, whereas Random Forest is better suited for complex, non-linear decision boundaries.

Four studies experimented with SVM \cite{algaolahi2021port, bertoli2021end, baah2022enhancing,  saikira2022detection}. All four also had explored RF performance. Results for the SVM experiments ranged from 89.61\% to 99.87\% both coming from the same dataset (of two total). Source code availability and port scan details remained the same as indicated in the RF algorithm category.

\begin{center}
\captionof{table}{Support Vector Machine Algorithm Performance}
\begin{tabular}{ccc}
	\textbf{Authors} & \textbf{Accuracy (F1)} & \textbf{Dataset} \\ \hline \hline
	Algaolahi et al. & 89.61 & CICIDS2017 \\
	Baah et al & 99.87 & CICIDS2017 \\ 
	SaiKiran et al. & 93.29 & CICIDS2017\\
	Bertoli et al. & 92.00 & Bonafide \\ \hline
\end{tabular}
\end{center}

\subsubsection{Regression}
Regression algorithms are used for predicting a continuous target variable based on one or more input features. They are commonly used for tasks such as predictions and forecasting.

Regression algorithms, including linear regression, are different from Random Forest and SVM algorithms in that they focus on establishing a linear or non-linear relationship between the input features and the target variable. On the other hand, Random Forest and SVM algorithms are mainly used for classification problems.

In a regression problem, the aim is to predict a numerical output, whereas in classification the output is categorical. SVM can also be used for regression problems by using a specific formulation called Support Vector Regression (SVR). However, the emphasis and method used in regression algorithms are different compared to SVM and Random Forest.

Four different datasets were used by three studies \cite{sirisha2021intrusion, bertoli2021end, Tojeiro2022}. The results span 59.21\% to 94\% accuracy (F1). Only one study \cite{bertoli2021end} discussed port scanning in detail and included source code for the algorithm. 
 
\begin{center}
\captionof{table}{Regression Algorithm Performance}
\begin{tabular}{ccc}
	\textbf{Authors} & \textbf{Accuracy (F1)} & \textbf{Dataset} \\ \hline \hline
	Tojeiro et al. & 94.00 & CICIDS2017 \\
	Sirisha et al. & 74.05 & NSLKDD \\
	Sirisha et al. & 59.21 & CICIDS2017 \\
	Bertoli et al. & 70.00 & MAWILab \\ 
	Bertoli et al. & 92.00 & Bonafide \\  \hline
\end{tabular}
\end{center}

\subsubsection{Naive Bayes}
Naive Bayes is a probabilistic algorithm that is good for classification problems, especially when the assumption of independence between features holds. It is fast and simple to implement and can handle large datasets well.

Naive Bayes is different from regression, SVM, and Random Forest algorithms in that it makes a probabilistic prediction based on Bayes' theorem and the assumption of independence between features, whereas the other algorithms make predictions based on a boundary or a combination of trees. In comparison, regression, SVM, and Random Forest algorithms work well for more complex problems where the relationship between features is not necessarily independent and the decision boundary is not clear.

Bakaletz \cite{bakaletz2022machine} used a custom generated dataset for training and evaluation. The author used two nmap scan techniques- aggressive (NMAP-A) and stealth (NMAP-S). There were five additional datasets used in three \cite{al2021detecting, sirisha2021intrusion, bertoli2021end} out of the remaining four studies \cite{baah2022enhancing} in this category. Training and evaluation of NB algorithms demonstrated performances in the range of 55\% to 99.7\%.  

\begin{center}
\captionof{table}{Naive Bayes Algorithm Performance}
\begin{tabular}{ccc}
	\textbf{Authors} & \textbf{Accuracy (F1)} & \textbf{Dataset} \\ \hline \hline
	Al-Haija et al. & 99.70 & PSA-2017\\
	Baah et al & 93.02 & CICIDS2017 \\ 
	Bakaletz & 98.00 & NMAP-A \\
	Bakaletz & 82.00 & NMAP-S \\
	Sirisha et al. & 74.47 & NSLKDD \\
	Sirisha et al. & 37.92 & CICIDS2017 \\
	Bertoli et al. & 55.00 & MAWILab\\ 
	Bertoli et al. & 78.00 & Bonafide \\ \hline
\end{tabular}
\end{center}

\subsubsection{Neural networks} 
Neural networks (NNs) are good for a wide range of tasks including classification, speech recognition, and natural language processing. They are particularly good for complex and non-linear problems, such as recognizing patterns where traditional algorithms struggle.

Neural networks are different from other machine learning algorithms in that they are based on a modeled structure inspired by the human brain. They consist of multiple interconnected nodes, or artificial neurons, that process information. These neurons are organized into layers and the connections between them are associated with weights that are learned during the training process.

In comparison, algorithms such as RF, SVM, NB, and regression algorithms make predictions based on a clear boundary or a combination of trees or linear relationships, whereas NNs are capable of learning complex relationships between the inputs and outputs through their internal structure. NN algorithms have the ability to learn and make predictions based on the examples they are trained on, which makes them highly flexible. However, they can also be more difficult to interpret and train, and require a large amount of data to achieve good performance.

Furthermore, we differentiate between two types of NN: Artificial Neural Networks (ANNs) and Convolutional Neural Networks (CNNs). Both types of deep learning algorithms used for various tasks. The main difference between ANNs and CNNs is the structure and the way they process information. ANNs are fully connected networks, where each neuron in one layer is connected to all neurons in the next layer. This structure makes ANNs computationally expensive and require a large amount of data to achieve meaningful results. 

On the other hand, CNNs are specifically designed for classification tasks. They have a unique structure, consisting of convolutional layers, activation layers, pooling layers, and fully connected layers. Convolutional layers are used to extract features, activation layers apply a non-linear activation function to the output of the convolutional layer, pooling layers reduce the spatial dimensions of the output, and fully connected layers make the final prediction. This unique structure makes CNNs efficient and effective for classification tasks, as it allows them to learn hierarchical representations of the data and identify different objects and features. In comparison, ANNs are not specifically designed for classification and may not achieve the same level of performance as CNNs for these types of tasks.

We did break our own time bounding constraint in this category because Aamir et al. \cite{Aamir2021} specifically noted neural networks were not being investigated. Hartpence et al. \cite{Hartpence2020} published their work in 2020, therefore we felt obliged to include the work here. 

On that note, three of the four studies in this algorithmic category used non-standard datasets. Hartpence et al. \cite{Hartpence2020} provided immense detail in what port scan types exist in the datasets the authors generated as part of their experiments. The general (GEN) dataset contained typical network traffic with port scans intermingled. The second dataset contained port scans only (TCP). Lv et al. \cite{lv2022application} generated a custom dataset by capturing network traffic while executing nmap full connect scans (NMAP-F). Kirtas et al. \cite{kirtas2022early} executed nmap SYN scans (NMAP-Y) while capturing network traffic for the dataset. No study included code snippets or links to code repositories. 

\begin{center}
\captionof{table}{ANN Algorithm Performance}
\begin{tabular}{ccc}
	\textbf{Authors} & \textbf{Accuracy (F1)} & \textbf{Dataset} \\ \hline \hline
	Hartpence et al. & 99.99 & GEN \\
	Hartpence et al. & 99.31 & TCP \\
	SaiKiran et al. & 99.11 & CICIDS2017 \\
	Kirtas et al. & 87.88 & NMAP-Y \\
	Bertoli et al. & 100.00 & Bonafide \\ \hline
\end{tabular}
\end{center}

\begin{center}
\captionof{table}{CNN Algorithm Performance}
\begin{tabular}{ccc}
	\textbf{Authors} & \textbf{Accuracy (F1)} & \textbf{Dataset} \\ \hline \hline
	Lv et al. & 99.00 & NMAP-F\\
	SaiKiran et al. & 63.52 & CICIDS2017 \\
	Singh et al. & 99.94 & CICIDS2017 \\
	Henry et al. & 98.73 & CICIDS2017 \\
\end{tabular}
\end{center}

\section{Conclusion}
We posed four research questions to guide this systematic review of port scan detection literature. We discovered five algorithms present in the literature: Random Forest, Support Vector Machine, Regression, Naive Bayes, and Neural Network. The literature revealed multiple ways to accurately detect port scanning. Bertoli et al. \cite{bertoli2021end} confirmed RF and ANN algorithms are capable of 100\% accuracy against a plethora of port scan types and techniques. Sirisha et al. \cite{sirisha2021intrusion} showed the poorest accuracy, 37.92\%, with the authors evaluation of the NB algorithm. ANN seemed to be the strongest type of algorithm across all of the sample papers, followed by RF. There were 11 different datasets used in 34 algorithm experiments with the CICIDS2017 dataset being used in 47\% of those experiments. These results originated from 14 research studies published since 2021 and a single study coming from 2020.  Unfortunately, we were unable to adequately address the fourth research question (i.e., \textit{What port scanning types and techniques were used for evaluation of those algorithms?}) as only a few studies discussed port scanning in any detail.

Overall, we found a variety of existing work included port scanning as a minor point compared to focuses areas such as general intrusion detection, DDoS, malware, botnets, and so forth. We assume any work demonstrating a capability to detect port scanning mentioned such explicitly and thus would be discoverable in our search. Another assumption underlying this work is research indexing. More specifically, we assume existing work has been indexed and thus was discoverable given our search strings. A final assumption present throughout the literature seems to be the stability of port scanning types and techniques. The dominance of the CICIDS2017 dataset in training and evaluation supports this point. This assumption will continue to be reasonable as long as significant innovations do not occur in the port scanning research.

It is interesting to note many existing papers experiment with more than one machine learning algorithm. The diversity of results within an algorithm category, across the sample papers in the category, is curious. Perhaps this would not be so unexpected if every paper used a different dataset for training and evaluation. Yet, much of the existing work leverages the CICIDS2017 data. As well, The distinct shift to NN algorithms in 2022 is notable. The prominence of NN over the past year suggests NN and its variations represent a viable research pathway going forward. As a final note, we found the vast majority of studies do not include code snippets or links to GitHub repositories.  

Accordingly, replication and reproduction to include specific port scanning techniques with packet captures and algorithm source code would be beneficial. This would be significant because doing so would fill in an existing gap and also enable adjacent research in cybersecurity such as offensive cyber, network security, and so forth. At the same time, such future work might look to incorporate  green compute measurements given the increased focus on sustainability in algorithm research. Foundational compute resource metrics can be taken from Bertoli et al. \cite{bertoli2021end} who detailed the compute resource profiles associated with training and evaluating the ML algorithms in their work.

A final area for future exploration is \textit{nature inspired} or \textit{artificial life} (Alife) algorithms. Greensmith et al. \cite{Greensmith2010} showed how a dendritic cell algorithm can be used to detect port scanning. The authors show a theoretical model with pseudocode examples. More recently, Liu et al. \cite{Liu2021} extended the dendritic cell concept albeit without evaluation against port scan datasets. Such algorithms should be evaluated using the datasets utilized in the review sample papers and additional Alife algorithms investigated.

\bibliographystyle{numeric}
\bibliography{references}

\end{multicols}
\end{document}